\documentclass[preprint,
superscriptaddress,
showpacs,
 amsmath,amssymb,
]{revtex4-1}

\usepackage{graphicx}% Include figure files
\usepackage{dcolumn}% Align table columns on decimal point
\usepackage{bm}% bold math
\usepackage{color}
\usepackage{ulem}
\begin{document}

%\preprint{APS/123-QED}

\title{Acceleration and deceleration of quantum dynamics based on inter-trajectory travel with fast-forward scaling theory}% Force line breaks with \\

\author{Shumpei Masuda}
\affiliation{Research Center for Emerging Computing Technologies, National Institute of Advanced Industrial Science and Technology (AIST), 1-1-1, Umezono, Tsukuba, Ibaraki 305-8568, Japan}
\email{shumpei.masuda@aist.go.jp}
\author{Jacob Koenig}
\affiliation{Kavli Institute of Nanoscience, Delft University of Technology, 
\\
Lorentzweg 1, 2628 CJ, Delft, The Netherlands}
\email{j.d.koenig@tudelft.nl}
\author{Gary A. Steele}
\affiliation{Kavli Institute of Nanoscience, Delft University of Technology, 
\\
Lorentzweg 1, 2628 CJ, Delft, The Netherlands}

%\affiliation{*masulas@tmd.ac.jp}

\date{\today}% It is always \today, today,

%\pacs{02.30.Yy, 37.90.+, 67.85.Fg, 03.75.Lm}% PACS, the Physics and Astronomy
\begin{abstract}
Quantum information processing requires fast manipulations of quantum systems in order to overcome dissipative effects. We propose a method to accelerate quantum dynamics and obtain a target state in a shorter time relative to  unmodified dynamics, and apply the theory to a system consisting of two linearly coupled qubits. We extend the technique to accelerate quantum adiabatic evolution in order to rapidly generate a desired target state, thereby realizing a shortcut to adiabaticity. Further, we address experimental limitations to the rate of change of control parameters for quantum devices which often limit one's ability to generate a desired target state with high fidelity. We show that an initial state following decelerated dynamics can reach a target state while varying control parameters more slowly, enabling more experimentally feasible driving schemes.

%\textcolor{blue}{[Comment 1] We propose a method to accelerate and decelerate quantum dynamics, and apply the theory to a system consisting of two linearly coupled qubits. 
%We extend the technique to accelerate quantum adiabatic evolution in order to rapidly generate a target state, thereby realizing a shortcut to adiabaticity.
%Quantum information processing requires fast manipulations of quantum systems. 
%Acceleration of quantum dynamics will provide a way to overcome decoherence effects in state preparations. 
%There are often experimental limitations to the rate of change of control parameters of a given system under examination.
%Simple slowing of control parameters generates different state because there is no simple scaling in quantum dynamics in general. 
%We show that the method for deceleration  can be used to find slower control parameters which generate approximately the same target state.}
\end{abstract}

\maketitle

{\bf Introduction--}
An essential ingredient to the further development of quantum technologies is the ability to rapidly and accurately control quantum systems in order to overcome the effects of decoherence. However, modification of the speed of quantum dynamics is often nontrivial in general due to both the lack of a simple scaling property in the dynamics as well as the infinitely large parameter spaces which one must generally navigate \cite{Masuda2008}. Thus, both experimentally feasible and non-trivial scaling properties in quantum dynamics are highly desirable to simplify the controls which regulate the time evolution of quantum systems.
%\textcolor{blue}{It is very difficult to find suitable time dependences of control parameters for speed control because there is an infinite number of choices.  }

Fast-forward scaling theory (FFST) provides a systematic way for optimally designing control parameters which accelerate, decelerate, or reverse the dynamics of a quantum system \cite{Masuda2008,MasudaRice2016}.
The formalism of FFST has previously been extended with great effect to many-body~\cite{Masuda2012} and discrete  systems ~\cite{Masuda2014,Takahashi2014,Xi2021}, systems of charged particles~\cite{Masuda2011,Setiawan2021}, tunneling dynamics~\cite{Khujakulov2016,Nakamura2017}, Dirac dynamics \cite{Deffner2015,Deffner2021} and for the acceleration of adiabatic dynamics~\cite{Masuda2010,Torrontegui2012,Patra2021}.
The application of FFST to adiabatic dynamics can produce what are known as shortcuts to adiabaticity (STA) or assisted adiabatic transformations \cite{Masuda2010,Torrontegui2012,MasudaRice2016}.
Protocols utilizing FFST with quantum, classical, and stochastic dynamics have also been previously proposed \cite{Patra2017,Jarzynski2017}.

When utilizing FFST, one can often obtain viable trajectories through the system's state space which realize the user's desired end state. However, as we will show in this paper, FFST is not applicable in some parameter regimes due to the lack of a suitable speed-controlled trajectory. Therefore, modification of the theory is required to resolve this issue. In this paper, we introduce a novel method which we call inter-trajectory travel (ITT), to resolve such deficiencies. 
Thus, our work addresses a fundamental challenge in quantum dynamics: the ability to control the rate of change of a quantum state. To demonstrate the effectiveness of ITT, we apply the framework to accelerate and decelerate the time evolution of two-level systems. Furthermore, we use ITT to realize shortcuts to adiabaticity by generating approximately the same state as that which is achieved by slower, adiabatic dynamics.

We focus in particular on deceleration in this study in contrast to previous works which have been largely concerned with fast and extremely precise controls.
Control parameters for fast and accurate state preparation often have rapidly varying time dependencies
when designed by other protocols \cite{Goerz2017,Larrouy2020}.
However, there are typically experimental limitations to the rate of change of control parameters of a given system under examination \cite{Sevriuk2019}.
%Actual form of control parameters can be distorted when the external fields are controlled very fast due to the imperfection of experimental setup even if source equipment such as arbitrary waveform generator can generate aimed form.
Naively scaling down the rate of change of control parameters will in general not produce the desired target state, leading to a loss of fidelity. 
Our method for deceleration can be used to find slower control parameters which reliably generate approximately the same target state in a longer time without iterative integration of the Schr\"{o}dinger equation.

In order to introduce our method we consider the acceleration and deceleration of a linearly coupled two-qubit system as an example. Although any arbitrary qubit state can be generated through a sequence of distinct single and two-qubit gates, it is often more convenient if one can generate a desired target state with fewer control parameters. In our method, the same single control parameter is modified with respect to the reference dynamics. Thus, our method does not require sophisticated manipulation of several control parameters, such as X, Y, and Z rotations of the qubits, but rather control over only the resonance frequency of a single qubit.

{\bf System--}
\label{Model}
In order to illustrate our method we consider a system of two coupled qubits as a concrete example, for which the Hamiltonian is represented as
\begin{eqnarray}
H/\hbar &=& \omega_1(t) \sigma_1^\dagger \sigma_1 + \omega_2 \sigma_2^\dagger \sigma_2
+ g(\sigma_1^\dagger \sigma_2 + \sigma_2^\dagger \sigma_1),
\label{H_qubit_10_4_20}
\end{eqnarray}
where $\omega_{i}$ and $g$ are the angular frequency of qubit $i$ and the coupling strength between the qubits, respectively.

Our system can be realized with a variety of platforms which enable frequency tunability of qubits. In particular, the field of circuit quantum electrodynamics~\cite{Gambetta2017,Wendin2017,Krantz2019,Gu2019,Blais2020} in which a superconducting qubit's transition frequency can be engineered to be modified by a magnetic flux threading its SQUID loop~\cite{Koch2007}, is a suitable candidate.
A realization of the system is discussed in Supplementary Section S1.%Appendix~\ref{superconducting qubits}.
%\textcolor{blue}{Refer to a paper which was introduced in APS.}

\begin{figure}
\begin{center}
\includegraphics[width=16cm]{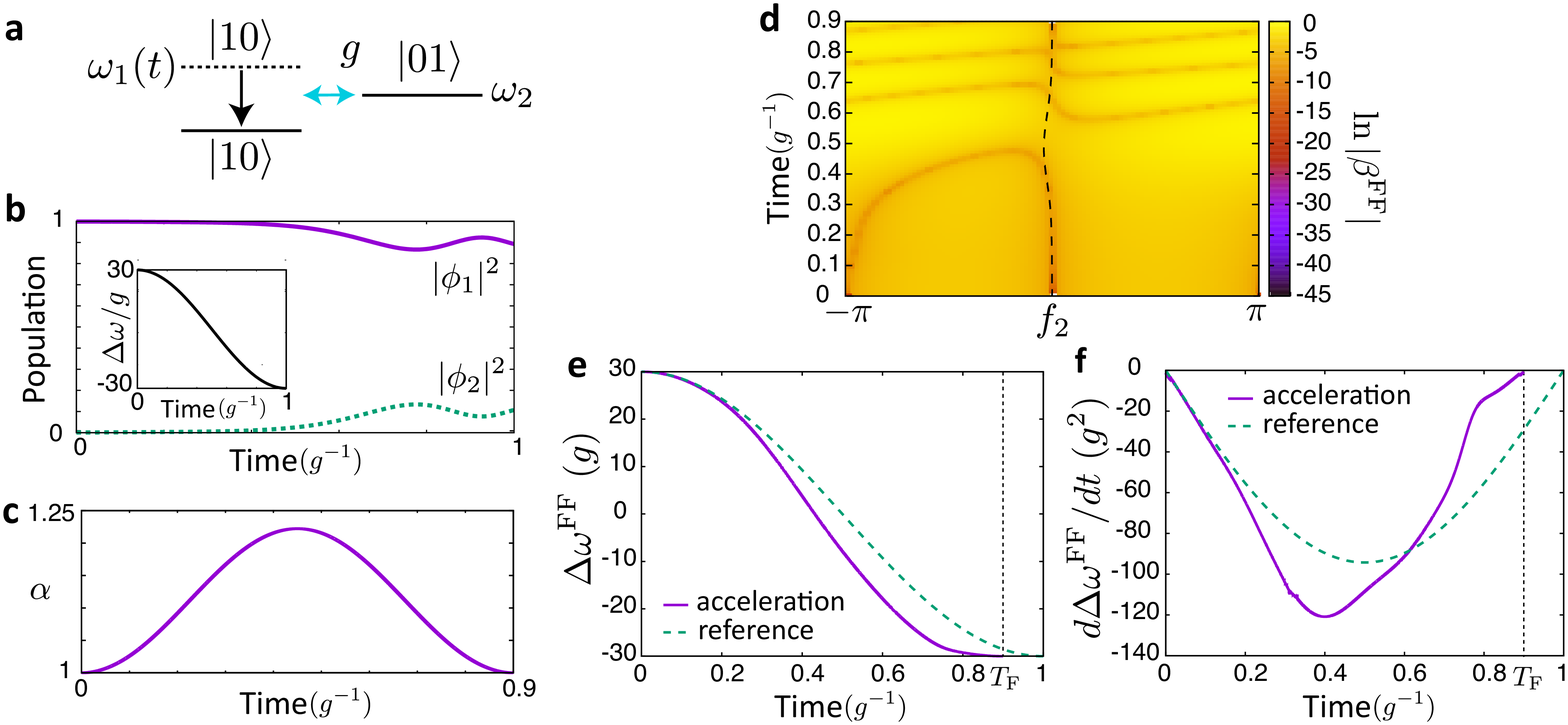}
\end{center}
\caption{
{\bf Schematic of the system, and speed-controlled and virtual trajectories for acceleration.}
{\bf a}, Schematic of the system. $\omega_1$ is decreased gradually, while $\omega_2$ and $g$ are fixed.
There is population transfer from $|10\rangle$ to $|01\rangle$.
{\bf b}, Time dependence of population of $|m\rangle$ in the reference dynamics. The inset shows the time dependence of $\Delta\omega=\omega_1-\omega_2$.
{\bf c}, Time dependence of the magnification factor $\alpha$ for the case of acceleration.
The used parameters are $\Delta\omega_{0}=30g$, $T=g^{-1}$ and $T_{\rm F}=0.9g^{-1}$.
{\bf d}, 
$\ln|\beta^{\rm FF}|$ as a function of $f_2$ and $t$ for $T_{\rm F}=0.9g^{-1}$.
The dashed curve shows a virtual trajectory.
{\bf e}, 
Time dependence of $\Delta\omega^{\rm FF}=\omega_1^{\rm FF}-\omega_2^{\rm FF}$ for the virtual trajectory and $\Delta\omega$ for the reference dynamics.
{\bf f},  
Time dependence of $d\Delta\omega^{\rm FF}/dt$ and $d\Delta\omega/dt$.
}
\label{system_10_4_20}
\end{figure}

We assume $\omega_1(t), \omega_2 \gg g$ for all times $t$.
We require that $\omega_1(0)-\omega_2 \gg g$ and that the initial state of the system is the energy eigenstate which is represented by $|10\rangle$, where the first and the second indices are for qubit 1 and qubit 2, respectively.
Then, $\omega_1(t)$ is decreased gradually, while $\omega_2$ and $g$ are fixed as illustrated in Fig.\ref{system_10_4_20}a.
As the qubit frequencies near resonance, there is finite population transfer from $|10\rangle$ to $|01\rangle$ due to the  coupling.

In the following analysis, we assume that  the total time evolution of the system occurs on a timescale far shorter than the relevant coherence times of the qubits.
Then, the dynamics of the system is confined to a subspace spanned by two states, $|{1}\rangle=|10\rangle$ and $|{2}\rangle=|01\rangle$. 
The state of the system under investigation is represented by
\begin{eqnarray}
|\Psi(t)\rangle = \phi_{1}(t)|1\rangle +  \phi_{2}(t)|2\rangle.
\end{eqnarray}
with the Schr\"{o}dinger equation written as
\begin{eqnarray}
i\frac{d}{dt}\phi_{m}(t) = g \phi_{l}(t) + \omega_m(t) \phi_{m}(t),
\label{SE_ref_10_24_20}
\end{eqnarray}
where hereafter $m,l\in\{1,2\}$ and $l\ne m$.

The dynamics of this system can be emulated also by a single qubit system under a drive after moving to a rotating frame and applying the rotating wave approximation (RWA) as explained in Supplementary Section S2. %Appendix~\ref{Single qubit}.

{\bf Fast-forward scaling theory--}
\label{Fast-forward scaling theory}
We derive the time dependence of $\omega_m$ which modifies the dynamics of the system, following the manner used in Ref.~\cite{Masuda2014}.
The reference dynamics which is to be accelerated or decelerated is defined by $\phi_m(t)$ which satisfies equation~(\ref{SE_ref_10_24_20}).
The  target state is defined as $\phi_m(T)$ for $T>0$.
We aim to generate the target state at a desired time $T_{\rm F}\ne T$ from a given initial state which is the same as in the reference dynamics.

We write the wave function of the speed-controlled dynamics in terms of 
the wave function of the reference dynamics, $\phi_m(t)$, as
\begin{eqnarray}
\phi_m^{\rm FF}(t) = \phi_m(\Lambda(t)) e^{if_m(t)},
\label{phiFF_1_31_21}
\end{eqnarray}
where $f_m(t)$ is an additional time-dependent phase and $\Lambda(t)$ is the scaled time defined by
\begin{eqnarray}
\Lambda(t) = \int_0^t \alpha(t')dt'.
\label{Lambda_2_26_21}
\end{eqnarray}
In equation~(\ref{phiFF_1_31_21}), $\phi_m(\Lambda(t))$ is the wave function of
the ideal dynamics naively scaled with respect to time. 
Here, $\alpha$ is called the {\it magnification factor}.
When $\alpha>1$, the time evolution of $\phi_m(\Lambda(t))$ is accelerated, while when $0<\alpha<1$ the  dynamics are slowed and  when $\alpha<0$, the dynamics are reversed. For instance, in the case where $\alpha=2$ 
the accelerated dynamics are twice as fast as the reference dynamics.
However $\phi_m(\Lambda(t))$ cannot be realized when $g$ is fixed. We introduce the additional phase $f_m(t)$ so that the state with the wave function in equation~(\ref{phiFF_1_31_21}) can be realized even with fixed $g$ 
(see Supplementary Section S3).
%(see Appendix \ref{Role of additional phase}).
The time dependence of $\alpha$ is chosen so that it satisfies
\begin{eqnarray}
\Lambda(T_{\rm F}) = T.
\end{eqnarray}
Note that the wave function satisfies $\phi_m^{\rm FF}(0) = \phi_m(0)$ and $\phi_m^{\rm FF}(T_{\rm F}) = \phi_m(T)$ if the additional phase vanishes at the initial and final time, $T_{\rm F}$.

We assume that $\phi_m^{\rm FF}$ is a solution of the Schr\"{o}dinger equation:
\begin{eqnarray}
i\frac{d}{dt}\phi_m^{\rm FF}(t) = g \phi_{l}^{\rm FF}(t) + \omega_m^{\rm FF}(t) \phi_m^{\rm FF}(t).
\label{SEFF_1_31_21}
\end{eqnarray}
The coupling strength is the same as in equation~(\ref{SE_ref_10_24_20}).
We substitute equation~(\ref{phiFF_1_31_21}) into equation~(\ref{SEFF_1_31_21}), divide by $\phi_m^{\rm FF}(t)$, and rearrange the equation into real and imaginary parts to obtain two equations:
\begin{eqnarray}
\alpha(t) {\rm Im} [ \phi_m^\ast(\Lambda(t)) \phi_{l}(\Lambda(t))] = {\rm Im}\{ \phi_m^\ast(\Lambda(t)) \phi_l(\Lambda(t)) \exp[i(f_l(t)-f_m(t))]\}
\label{f_10_10_20}
\end{eqnarray}
and  
\begin{eqnarray}
\omega_m^{\rm FF}(t) &=& {\rm Re} \Big\{g\frac{ \phi_l(\Lambda(t))}{\phi_m(\Lambda(t))} \Big{(}\alpha(t)-\exp[i(f_l(t)-f_m(t))]  \Big{)} \Big\}\nonumber\\
&& + \alpha(t) \omega_m(\Lambda(t)) - \frac{df_m(t)}{dt},
\label{V_10_10_20}
\end{eqnarray}
where $l\ne m$. Equation (\ref{f_10_10_20}) is used to obtain the additional phase $f_m(t)$, and
equation~(\ref{V_10_10_20}) is used to calculate the time-dependent qubit resonance frequency which yields the speed-controlled dynamics.
Note that $f_{m(l)}(t)=0$ is a solution of equation~(\ref{f_10_10_20}) when $\alpha=1$. We set $f_1(t)=0$ and consider variations in $f_2(t)$. This is justified given that only the phase difference $f_1(t)-f_2(t)$ is relevant for the dynamics.
The above formalism can also be extended to the case in which there exists a tunable coupling $g(t)$, as shown in 
Supplementary Section S4.
%Appendix~\ref{Tunable coupling}.

{\bf Acceleration--}
In order to generate the desired target state from a given initial state, one first must determine the additional phase which vanishes at the initial and final times.
However sometimes there exist no such solutions to equation~(\ref{f_10_10_20}).
Here, we develop the ITT method which realizes the target state approximately in cases where no exact solutions would ordinarily exist.

We consider the acceleration of some reference dynamics, in which
$\omega_1$ is decreased for $0\le t \le T$ as 
\begin{eqnarray}
\omega_1(t) =\Delta\omega_{0} \cos(\pi t/ T) + \omega_2,
\end{eqnarray}
while $\omega_2$ and $g$ are held constant.
The time dependence of $ \omega_1$ and the population of $|m\rangle$ are shown in Fig.~\ref{system_10_4_20}b.
The wave function of the reference dynamics $\phi_{m}(t)$ is obtained by solving the Schr\"{o}dinger equation~(\ref{SE_ref_10_24_20}) numerically. We consider the acceleration and deceleration of these particular
dynamics (the ``reference dynamics") in the following, as this is simply one such case where ITT resolves the shortcomings of FFST.

As an example, we use the magnification factor defined by
\begin{eqnarray}
\alpha(t) = 1 - \frac{T_{\rm F}-T}{T_{\rm F}}\{1 - \cos(2\pi t/T_{\rm F}) \},
\label{alpha_3_2_21}
\end{eqnarray}
where $\alpha$ is chosen to satisfy $\alpha(0) = \alpha(T_{\rm F})=1$ so that the speed-controlled dynamics coincides with the reference dynamics at $t=0$ and $T_{\rm F}$.
For $T_{\rm F}<T$ (acceleration), the magnification factor satisfies $\alpha \ge 1$.
The time dependence of $\alpha$ is shown in Fig.~\ref{system_10_4_20}c.
In this example, the accelerated dynamics takes 0.9 times less than the reference to reach the desired state.

Figure~\ref{system_10_4_20}d shows $\beta^{\rm FF}$ defined as
\begin{eqnarray}
\beta^{\rm FF}(t,f_2)/g= \alpha(t) {\rm Im} [ \phi_1^\ast(\Lambda(t)) \phi_{2}(\Lambda(t))] - {\rm Im}\{ \phi_1^\ast(\Lambda(t)) \phi_2(\Lambda(t)) \exp[if_2]\},
\label{beta_10_24_20}
\end{eqnarray}
which is the difference of the left hand side and the right hand side of equation~(\ref{f_10_10_20}) for $f_1(t)=0$. Note that $f_2$ is regarded as a variable in equation (\ref{beta_10_24_20}) instead of a solution of equation~(\ref{f_10_10_20}).
We plot $\beta^{\rm FF}(t,f_2)$ only for $-\pi < f_2 < \pi$ given that it is periodic with respect to $f_2$.
We note that the zeros of $\beta^{\rm FF}(t,f_2)$ correspond to the solutions of equation~(\ref{f_10_10_20}).

The trajectories, which are defined by the $f_2(t)$ which satisfy $\beta^{\rm FF}(t,f_2(t))=0$, represent the realizable accelerated dynamics.
We call these paths ``speed-controlled trajectories".
However, in this particular case, there exist no trajectories which can connect the initial state corresponding to $f_2(0)=0$ and the target state $f_2(T_{\rm F})=0$ given that
there are no zeroes of $\beta^{\rm FF}(t,f_2)$ around $t=$0.5$g^{-1}$, 0.7$g^{-1}$ and 0.8$g^{-1}$.
Thus, the dynamics cannot be accelerated exactly. 
The mechanism by which gaps between trajectories open for acceleration and deceleration is explained in 
Supplementary Section S5.
%Appendix~\ref{Gaps between trajectories}.

In order to resolve the lack of a continuous path between the initial and final states, we introduce 
virtual trajectories which allow for navigation across sufficiently shallow gaps.
We consider the virtual trajectory depicted in Fig.~\ref{system_10_4_20}d indicated by a dashed line.
The virtual trajectory satisfies $f_2(0)=f_2(T_{\rm F})=0$ and $\beta^{\rm FF}(t,f_2)\simeq 0$ for all times throughout the system's evolution.
$\omega_m^{\rm FF}$ can then be calculated for any given virtual trajectory by substituting the corresponding $f_2(t)$ and $f_1(t)=0$ into equation~(\ref{beta_10_24_20}).
While both $\omega_1^{\rm FF}$ and $\omega_2^{\rm FF}$ may be time-dependent in general, only the difference between the angular frequencies, $\Delta\omega^{\rm FF}=\omega_1^{\rm FF}-\omega_2^{\rm FF}$, is of physical importance in this subspace. Thus, only one qubit frequency is required to be tunable, yielding a change to the global phase of the wave function 
(see Supplementary Section S6).
%(see Appendix \ref{Relative phase}).

The time dependence of $\Delta\omega^{\rm FF}$ and its time derivative corresponding to this virtual trajectory are shown in Fig.~\ref{system_10_4_20}e and \ref{system_10_4_20}f, respectively.
%We could put the comparison of $\omega_m$ with the case of $f_2=0$.
We define the fidelity of the control by the overlap, $|\langle \Psi_{\rm ref} | \Psi_{\rm ITT} \rangle|$, between the end state $|\Psi_{\rm ITT} (T_{\rm F})\rangle$ of the control with $\omega_m^{\rm FF}$ and the end state $|\Psi_{\rm ref} (T)\rangle=\sum_m \phi_m(T)|m\rangle$ of the reference dynamics.
The fidelity of the control with ITT is 0.9996 while the fidelity of the control with the naively accelerated control parameters, $\omega_m(\Lambda(t))$, is 0.9871.
%\textcolor{blue}{of the control without ITT} corresponding to $f_2=0$ is \textcolor{red}{0.9871}.
%\textcolor{blue}{In the control with $f_2=0$, the time dependence  of  the qubit resonance frequency is simply scaled as $\omega_m(\Lambda(t))$.
Therefore, this result shows that ITT can significantly improve the control fidelity compared to a naive scaling of the control parameters with respect to time.

The control with $\alpha(t)\omega_m(\Lambda(t))$, which approximates $\omega^{\rm FF}_m$ in equation (\ref{V_10_10_20}), also improves the fidelity compared to the control with $\omega_m(\Lambda(t))$. The fidelity of the control in this case is 0.9989. The improvement of the fidelity for this case is alternatively interpreted as follows. As explained in 
%Appendix \ref{Tunable coupling}
Supplementary Section S4, the ideal dynamics straightforwardly scaled with respect to time is realized if both the coupling strength and angular frequency of the qubits are scaled as $g^{\rm FF}(t)=\alpha(t) g$, $\omega_m^{\rm FF}(t)=\alpha(t) \omega_m(\Lambda(t))$. The control with $\alpha(t)\omega_m(\Lambda(t))$ and a fixed coupling strength approximates such dynamics, and thus it also improves the fidelity relative to the naively accelerated control.

{\bf Inter-trajectory travel for shortcuts to adiabaticity--}
In this section, we show that ITT can be used to realize shortcuts to adiabaticity.
As an example, we consider the case that $\omega_1(t)$ in equation~(\ref{SE_ref_10_24_20}) is gradually changed while the other parameters are fixed. If $\omega_1(t)$ is changed slowly enough and the initial state is an eigenstate of an initial Hamiltonian,
the state remains in the corresponding instantaneous energy eigenstate of the time-dependent Hamiltonian throughout the system's evolution.
On the other hand, rapid changes in $\omega_1(t)$ cause undesired nonadiabatic population transfer to other energy eigenstates and thus increase infidelity.
It has been previously shown that FFST can exactly realize the same final state as is produced adiabatically in a time shorter than the adiabatic timescale. 
In our method, only the detuning is modified in contrast to other methods which require modulation of the coupling~\cite{Berry2009,Chen2010,Chen2011}.
However, FFST alone cannot be utilized due to the lack of a suitable trajectory when the manipulation time is too short. 
In the following, we show that ITT can greatly suppress nonadiabatic transitions in such regimes.

We consider some ideal dynamics for which the wave function may be written as
$\phi_m(\omega_1(t))e^{-\frac{i}{\hbar}\int_0^t E(\omega_1(t'))dt'}$, 
where $\phi_m(\omega_1)$ is the wave function of an instantaneous energy eigenstate which satisfies 
\begin{eqnarray}
g \phi_{l}(\omega_1) + \omega_m \phi_{m}(\omega_1) = \frac{E(\omega_1)}{\hbar} \phi_{m}(\omega_1),
\label{SE_ref_3_8_21}
\end{eqnarray}
where $E(\omega_1)$ is the eigenenergy, and again $m,l\in\{1,2\}$ and $l\ne m$. 
When $\omega_1(t)$ is changed slowly enough, this is a solution of the Schr\"{o}dinger equation (\ref{SE_ref_10_24_20}).
On the other hand, the state deviates from the expected dynamics when $\omega_1(t)$ is changed on short timescales. 
We aim at finding angular frequencies which drive the initial state, $\phi_m(\omega_1(0))$,
to the target state, $\phi_m(\omega_1(T_{\rm F}))e^{-\frac{i}{\hbar}\int_0^{T_{\rm F}} E(\omega_1(t'))dt'}$, in a short time $T_{\rm F}$.

In FFST one assumes that the wave function of the speed-controlled dynamics has the form
\begin{eqnarray}
\phi_m^{\rm FF}(t) = \phi_m(\omega_1(t)) e^{if_m(t)}e^{-\frac{i}{\hbar}\int_0^t E(\omega_1(t'))dt'}.
\label{phiFF_3_8_21}
\end{eqnarray}
We assume that $\phi_m^{\rm FF}$ also satisfies equation~(\ref{SEFF_1_31_21}).
Using equations~(\ref{SEFF_1_31_21}), (\ref{SE_ref_3_8_21}) and (\ref{phiFF_3_8_21}), we obtain 
\begin{eqnarray}
\frac{d\phi_m(\omega_1(t))}{dt} = g \phi_l (\omega_1(t)) \sin[f_l(t) - f_m(t)]
\label{f_sta_3_8_21}
\end{eqnarray}
and 
\begin{eqnarray}
\omega_m^{\rm FF}(t) &=& \omega_m(t) + g\frac{\phi_{l}(\omega_1(t))}{\phi_m(\omega_1(t))}\nonumber\\
&& \times \{ 1 - \cos[ f_{l}(t) - f_m(t)] \} - \frac{df_m(t)}{dt},
\label{omega_sta_3_8_21}
\end{eqnarray}
where again, $m,l\in\{1,2\}$ and $l\ne m$.
Equation~(\ref{f_sta_3_8_21}) is used to calculate the additional phase, $f_m(t)$, while
equation~(\ref{omega_sta_3_8_21}) is used to calculate the angular frequency, $\omega_m^{\rm FF}$.
We set $f_1(t)=0$ as we did in the previous subsection.
The time dependence of $\omega_1$ is given by
\begin{eqnarray}
\omega_1(t) = \Delta\omega_0 \cos(\pi t/T_{\rm F}) + \omega_2,
\label{omega_3_11_21}
\end{eqnarray}
where $\Delta\omega_0$ is constant, and $T_{\rm F}$ is the final time of the control.

\begin{figure}
\begin{center}
\includegraphics[width=14.0cm]{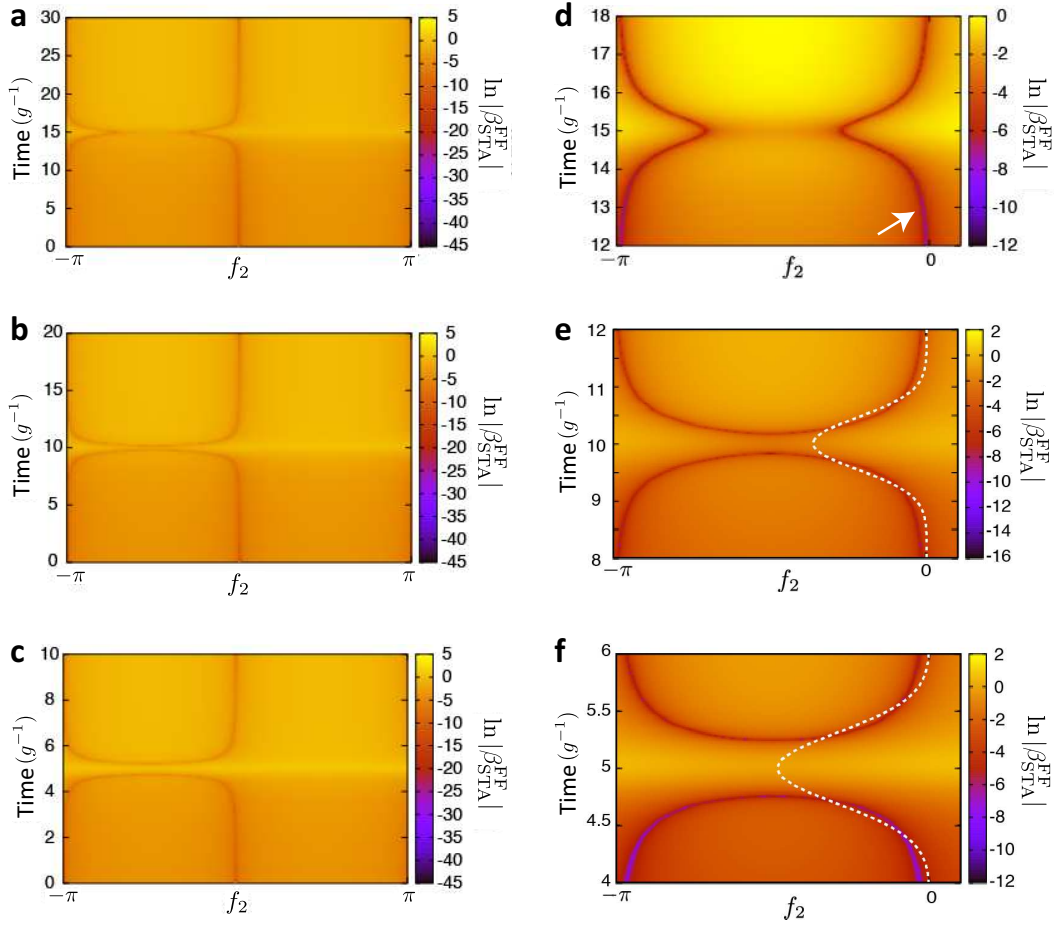}
\end{center}
\caption{
{\bf Inter-trajectory travel for shortcuts to adiabaticity.}
{\bf a,b,c,} $\ln|\beta^{\rm FF}_{\rm STA}|$ as a function of $f_2$ and $t$.
{\bf a}, {\bf b} and {\bf c} are for $T_{\rm F}=30g^{-1}$, $20g^{-1}$ and $10g^{-1}$, respectively.
{\bf d}, {\bf e} and {\bf f} are the closeups of panels {\bf a}, {\bf b} and {\bf c}, respectively.
The arrow in {\bf d} indicates the speed-controlled trajectory.
The white dashed curve in panels {\bf e} and {\bf f} shows a virtual trajectory.
$\Delta\omega_0=30g$, where $g$ is constant.
}
\label{theta_check_com_3_9_21}
\end{figure}

Figure~\ref{theta_check_com_3_9_21} shows the intensity of $\beta^{\rm FF}_{\rm STA}$ defined as
\begin{eqnarray}
\beta^{\rm FF}_{\rm STA}(t,f_2) = \frac{d\phi_1(\omega_1(t))}{dt} - g \phi_2 (\omega_1(t)) \sin[f_2]
\label{beta_STA_3_11_21}
\end{eqnarray}
which is the difference of the left hand side and the right hand side of equation~(\ref{f_sta_3_8_21}) for $f_1(t)=0$.
For a sufficiently long time $T_{\rm F}$, there exists a suitable trajectory which connects the initial state corresponding to $f_2(0)=0$ and the target state $f_2(T_{\rm F})=0$ as shown in Fig.~\ref{theta_check_com_3_9_21}a and \ref{theta_check_com_3_9_21}d.
$\omega_m^{\rm FF}(t)$ is obtained using equation~(\ref{omega_sta_3_8_21}), and $f_2(t)$, which satisfies 
$\beta^{\rm FF}_{\rm STA}(t,f_2)=0$, corresponds to the speed-controlled trajectory.
The obtained $\omega_m^{\rm FF}(t)$ can realize the target state exactly.
There are two trajectories because there are two values of $f_2(t)$ which satisfy equation~(\ref{f_sta_3_8_21}) in general.
One of the trajectories which satisfies $f_2(0)=f_2(T_{\rm F})=0$ is used to realize the STA.
The other trajectory generates different dynamics given a different initial state.

When $T_{\rm F}$ is not sufficiently long, there is no suitable trajectory for the fast-forward protocol as shown in Figs.~\ref{theta_check_com_3_9_21}{b,c} and \ref{theta_check_com_3_9_21}{e,f}.
The vertical gap around $t=T_{\rm F}/2$ in Figs.~\ref{theta_check_com_3_9_21}{b,c,e,f} is due to the lack of a solution for equation~(\ref{f_sta_3_8_21}).
We introduce a virtual trajectory which interconnects the two trajectories satisfying $f_2(0)=f_2(T_{\rm F})=0$ as represented 
in Figs.~\ref{theta_check_com_3_9_21}{e,f}.
The time-dependent frequency $\omega_m^{\rm FF}(t)$ calculated with these virtual trajectories can realize the target state approximately.
We use a Gaussian form for the virtual trajectory, and determine the parameters of the curve such that $\int_0^{T_{\rm F}} |\beta_{\rm STA}^{\rm FF}(t,f_2(t))|dt$ is minimized. 

\begin{figure}
\begin{center}
\includegraphics[width=14.0cm]{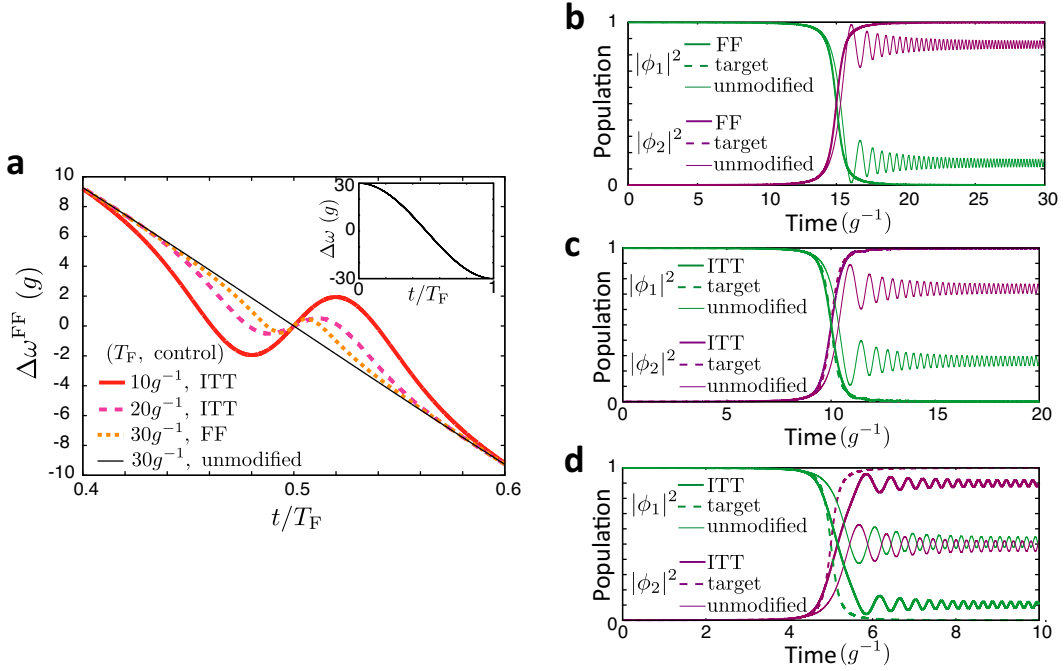}
\end{center}
\caption{
{\bf Time dependence of qubit resonance frequency and populations.}
{\bf a}, Time dependence of $\Delta\omega$ for the unmodified control with $T_{\rm F}=30g^{-1}$ (thin black solid curve), $\Delta\omega^{\rm FF}$ for FFST with $T_{\rm F}=30g^{-1}$ (orange dotted curve),
$\Delta\omega^{\rm FF}$ for ITT with $T_{\rm F}=20g^{-1}$ (pink dashed curve) and $T_{\rm F}=10g^{-1}$ (red solid curve).
{\bf b}, Time dependence of the population of $|m\rangle$ in the unmodified control and the control with FFST for $T_{\rm F}=30g^{-1}$. 
The corresponding speed-controlled trajectory is shown in Fig.~\ref{theta_check_com_3_9_21}a and \ref{theta_check_com_3_9_21}d.
The population of $|m\rangle$ in the target dynamics, $|\phi_m(\omega_m(t))|^2$, is also shown.
The curves for the target dynamics are overlapping with the ones for FFST.
{\bf c,d}, Time dependence of the population of $|m\rangle$ for the unmodified control and ITT in Fig.~\ref{theta_check_com_3_9_21}e and \ref{theta_check_com_3_9_21}f.
The populations in the target dynamics are also shown.
}
\label{omegaFF_com2_3_11_21}
\end{figure}

We compare the results of FFST and ITT with the unmodified control which utilizes the unmodified angular frequency, $\omega_1(t)$, in equation (\ref{omega_3_11_21}).
Figure~\ref{omegaFF_com2_3_11_21}a shows the time dependence of the difference between the angular frequencies for the unmodified control, FFST, and ITT.
It is seen that the angular frequencies are most drastically adjusted at the halfway point of evolution
around $t=T_{\rm F}/2$ when the wave function radically changes. 
The modification becomes larger as $T_{\rm F}$ is made shorter corresponding to a widening of the gap between trajectories.

Figure~\ref{omegaFF_com2_3_11_21}{b--d} show the time dependence of the population of $|m\rangle$ for the 
unmodified control, the controls with FFST, and with ITT.
In the target dynamics, $|\phi_1|^2\simeq 0$ and $|\phi_2|^2\simeq 1$ at $t=T_{\rm F}$. 
On the other hand, $|\phi_1|^2$ and $|\phi_2|^2$ deviate from their desired populations at $t=T_{\rm F}$
for the dynamics defined by the trajectory where $f_2(t)=0$ at all times, due to unwanted nonadiabatic effects.
Figure~\ref{omegaFF_com2_3_11_21}b shows that FFST realizes the exact target dynamics, while the fidelity, which is defined by the overlap with the target state at $t=T_{\rm F}$, for the unmodified control is 0.929.
Figure~\ref{omegaFF_com2_3_11_21}c and \ref{omegaFF_com2_3_11_21}d show that ITT can suppress nonadiabatic contributions and faithfully realize the approximate target state.
The fidelities of the controls are 0.999 and 0.949 for $T_{\rm F}=20g^{-1}$ and $T_{\rm F}=10g^{-1}$ with ITT, while the fidelities are  0.857 and 0.697 for the unmodified controls with $T_{\rm F}=20g^{-1}$ and $T_{\rm F}=10g^{-1}$, respectively. 
The fidelity when utilizing ITT is considerably higher than that of the unmodified controls, although the efficiency of ITT is also degraded as $T_{\rm F}$ becomes shorter due to the gap between the speed-controlled trajectories widening as seen in Fig.~\ref{theta_check_com_3_9_21}f.

{\bf Deceleration based on inter-trajectory travel--}
We next consider deceleration of the reference dynamics based on ITT.
We use the same form of the magnification factor, $\alpha(t)$, as equation~(\ref{alpha_3_2_21}) with  $T_{\rm F}>T$ such that $0 < \alpha(t) \le 1$ is satisfied for the decelerated dynamics.

\begin{figure}
\begin{center}
\includegraphics[width=12cm]{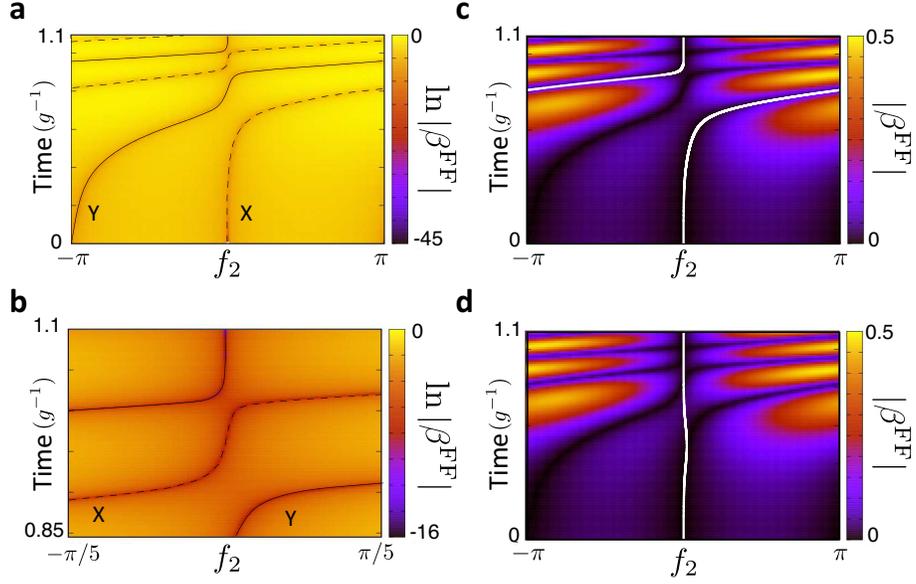}
\end{center}
\caption{
{\bf Speed-controlled and virtual trajectories for deceleration.}
{\bf a}, $\ln|\beta^{\rm FF}|$ as a function of $f_2$ and $t$ for $T_{\rm F}=1.1g^{-1}$.
Other parameters used are the same as in Fig.~\ref{system_10_4_20}b.
The dashed and solid curves indicate SCT-X and SCT-Y.
The dashed line does not reach $f_2=0$ at $t=1.1g^{-1}$.
{\bf b}, A closeup of {\bf a}. 
{\bf c},  VT-A indicated by the white curve interpolating the speed-controlled trajectories at around $t=0.9g^{-1}$.
{\bf d},  VT-B interpolating the speed-controlled trajectories at around $t=0.7g^{-1}$ and 0.9$g^{-1}$.
The color in {\bf c,d} shows $|\beta^{\rm FF}|$ as a function of $f_2$ and t.
}
\label{figure4_6_22_21}
\end{figure}

Figure~\ref{figure4_6_22_21}a,b shows $\beta^{\rm FF}$ as a function of $f_2$ and $t$ for $f_1=0$ and $T_{\rm F}=1.1g^{-1}$.
In this example, the decelerated dynamics takes 1.1 times longer than the unmodified dynamics to reach the desired state.
For the parameters we consider, there are two speed-controlled trajectories, X and Y (SCT-X and  SCT-Y), as represented in Fig.~\ref{figure4_6_22_21}a,b given that there are two possible sets of values for $f_2(t)$ which satisfy $\beta^{\rm FF}(t,f_2)=0$.
As seen in Fig.~\ref{figure4_6_22_21}a,b, there are narrow gaps between the speed-controlled trajectories
around $t=0.7g^{-1},~0.9g^{-1}$ and $g^{-1}$.
Importantly, there exist no trajectories which can connect the initial state corresponding to $f_2(0)=0$ and the target state $f_2(T_{\rm F})=0$. 

As shown previously for the case of accelerated dynamics, ITT can also approximately realize the desired end state for decelerated dynamics.
We consider two of the possible virtual trajectories in this study.
The virtual trajectories are shown in Fig.~\ref{figure4_6_22_21}c (virtual trajectory A [VT-A]) and Fig.~\ref{figure4_6_22_21}d (virtual trajectory B [VT-B]).
They satisfy $f_2(0)=f_2(T_{\rm F})=0$ and $\beta^{\rm FF}(t,f_2(t))\simeq 0$ for all times.
$f_2=\pi$ and $-\pi$ are regarded as the same point given that $\beta^{\rm FF}(t,f_2)$ is periodic with respect to $f_2$. 
Thus, VT-A is also continuous, although there is a jump from $\pi$ to $-\pi$ in Fig.~\ref{figure4_6_22_21}c.
We show in the following that the state of the system can approximately trace a selected virtual trajectory, although the virtual trajectory is not an exact solution of the Schr\"{o}dinger equation.

\begin{figure}
\begin{center}
\includegraphics[width=15.0cm]{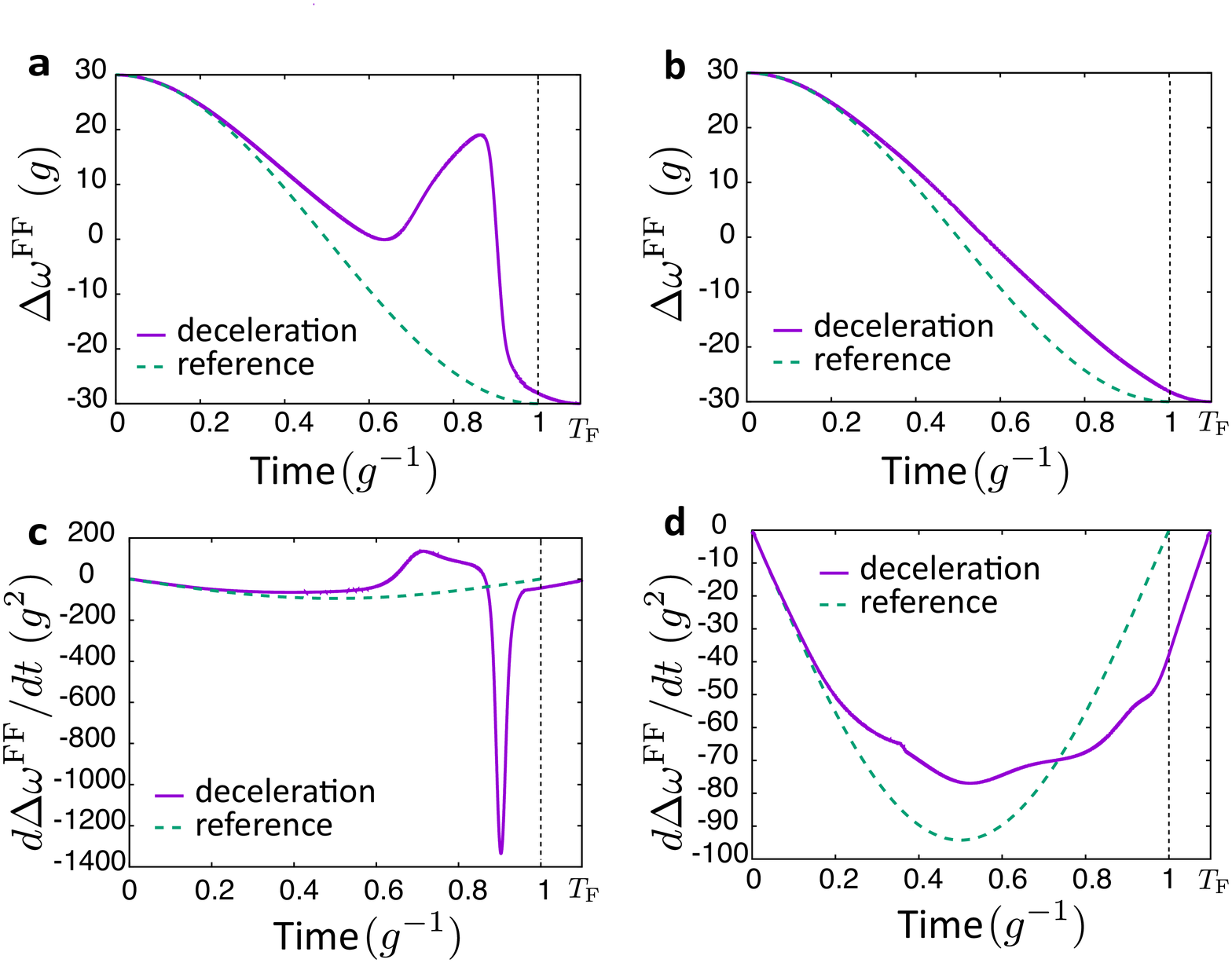}
\end{center}
\caption{
{\bf Difference of qubit frequencies and its time derivative.} {\bf a,b,} Time dependence of $\Delta\omega^{\rm FF}=\omega_1^{\rm FF}-\omega_2^{\rm FF}$ for the virtual trajectories. 
The dashed curve represents the $\Delta\omega=\omega_1-\omega_2$ used in the reference dynamics.
The parameters used are the same as in Fig.~\ref{figure4_6_22_21}.
{\bf c,d,} Time dependence of $d\Delta\omega^{\rm FF}/dt$ and $d\Delta\omega/dt$. 
{\bf a,c} are for virtual trajectory A, and {\bf b,d} are for virtual trajectory B.
}
\label{VFF_10_25_20}
\end{figure}

Figure~\ref{VFF_10_25_20} shows the time dependence of $\Delta\omega^{\rm FF}$ and its time derivative for both trajectories.
The time dependence of $\Delta\omega^{\rm FF}$ for VT-A is complicated compared to the one for VT-B due to the rate of change of $f_2(t)$ for VT-A being more rapid than for VT-B.
The fidelity of the control is 0.99998 and 0.9995 for VT-A and VT-B, respectively, while the fidelity of the control with the naively decelerated control parameters, $\omega_m(\Lambda(t))$, is 0.9876. 
The fidelity of the control with $\alpha(t)\omega_m(\Lambda(t))$ is 0.9984.

\begin{figure}
\begin{center}
\includegraphics[width=15.0cm]{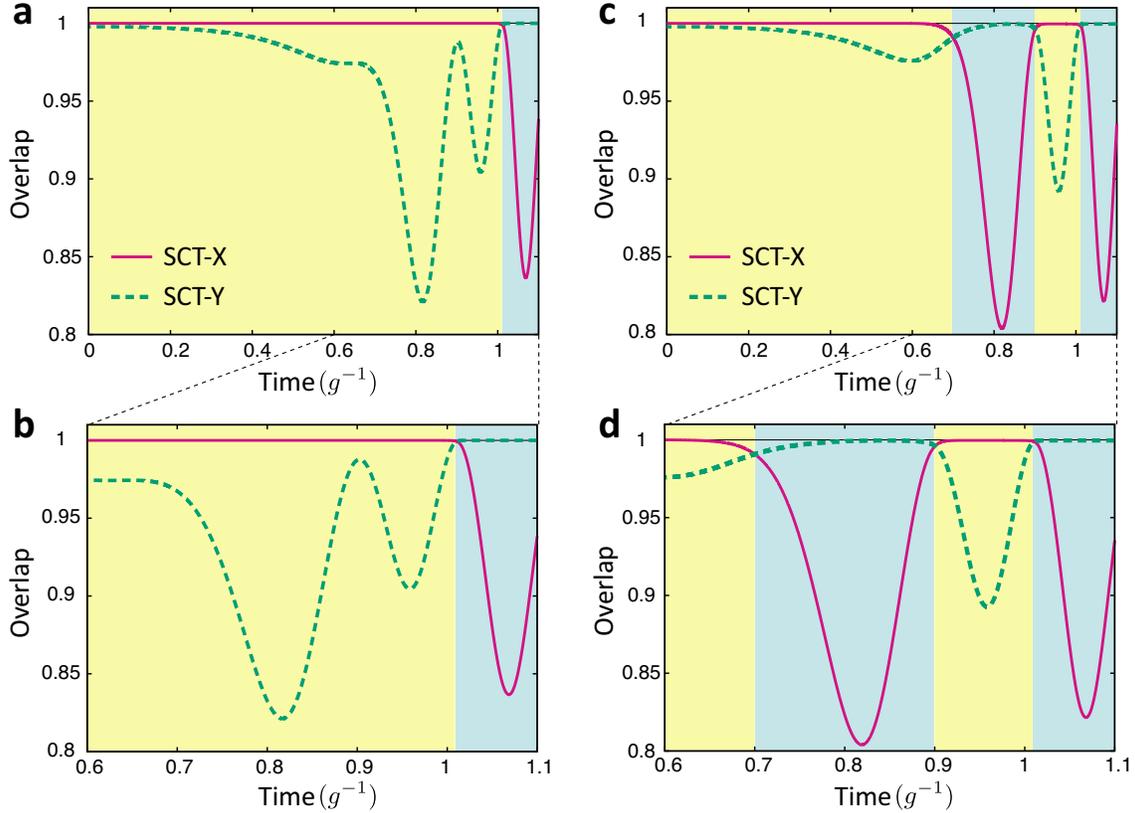}
\end{center}
\caption{
{\bf Overlaps with the speed-controlled trajectories.} 
{\bf a,b} are for ITT A and {\bf c,d} are for ITT B.
In the yellow region, the overlap with the SCT-X is larger than the one with the SCT-Y, while the overlap with the SCT-Y is larger than the one with the SCT-X in the blue region.
}
\label{fidelity_trajectories_3_12_21}
\end{figure}

Figure~\ref{fidelity_trajectories_3_12_21} illustrates the shifts between viable speed-controlled trajectories $|\Psi^{\rm FF}_{X/Y}(t)\rangle=\sum_m\phi_{m,X/Y}^{\rm FF}(t)|m\rangle$ that occur while a state follows a virtual trajectory given by $|\Psi_{\rm ITT} (t)\rangle$, where $\phi_{m,X/Y}^{\rm FF}(t)$ is defined by equation (\ref{phiFF_1_31_21}) with $f_m(t)$ corresponding to each speed-controlled trajectory. VT-A initially starts from SCT-X and approximately traces it, and near the end of its evolution shifts to SCT-Y as shown in Fig.~\ref{fidelity_trajectories_3_12_21}a. In the yellow region, the overlap with the trajectory X is greater than with the trajectory Y. In the light blue region, the overlap with trajectory Y is dominant. Thus, this result indicates the occurrence of trajectory shifts. An ITT event occurs once for VT-A and three times for VT-B as shown in Fig.~\ref{fidelity_trajectories_3_12_21}a and Fig.~\ref{fidelity_trajectories_3_12_21}b, respectively.

{\bf Conclusions--}
We have developed a novel method for the acceleration and deceleration of quantum dynamics, which we call inter-trajectory travel (ITT).
ITT is based on the knowledge of the structure of speed-controlled trajectories and gaps between those trajectories.
A virtual trajectory interconnecting different speed-controlled trajectories enables one to derive control parameters which either accelerate or decelerate the dynamics of a quantum system. 
Our method has extended the applicability of FFST by overcoming the non-existence of viable trajectories in the existing theory.
Furthermore, we have applied ITT to the study of shortcuts to adiabaticity and successfully shown that the same target state can be realized in a shorter time when compared to the adiabatic dynamics by suppressing unwanted nonadiabatic transitions.
The acceleration of quantum dynamics via ITT provides a novel way to outrun decoherence effects when manipulating quantum dynamics by solely modifying qubit frequencies.

We have also shown that the application of ITT for deceleration can be used to find slower control parameters which generate approximately the same target state.
We consider ITT to be useful for state preparation with modern quantum technologies as it allows one to design control parameters so that they may satisfy experimental limitations in laboratory
control hardware by loosening the often strict requirement of rapid and precise variation of
parameters.

An advantage of ITT is that it does not require iterative integrations of equations of motion in contrast to trial \& error protocols such as quantum optimal control theories.
Importantly, our method is complementary with other protocols.
For example, our method can be used to modify the speed of the dynamics derived by other protocols in order to
make the control parameters more experimentally feasible or to make the control duration shorter.
Thus, our technique adds to the quantum control toolbox which experimentalists may draw from to determine optimal parameters \cite{Motzoi2009, Motzoi2011, Gambetta2011, Schutjens2013, MartinezGaraot2015, Theis2016, Theis2018}.

\vspace{0.4cm}

{\bf Acknowledgements}\\
S.M. acknowledges the support from JST [Moonshot R \& D] [Grant Number JPMJMS2061] and JSPS KAKENHI [Grant Number 18K03486].  J.K. acknowledges support from the European Union’s Horizon 2020 research and innovation programme under grant agreement No 828826 (FET-Open Quromorphic).
The authors thank A. Sanz Mora for fruitful discussions.

{\bf Author contributions}\\
S.M. and J.K. carried out the theoretical analysis and numerical simulations, and wrote the manuscript with input from G.S. All authors reviewed the manuscript.

{\bf Competing interests}\\
The authors declare no competing interests.

%\\

\clearpage

\setcounter{figure}{0}
\renewcommand*{\thefigure}{S\arabic{figure}}
\setcounter{table}{0}
\renewcommand*{\thetable}{S\arabic{table}}
\setcounter{page}{1}
\setcounter{equation}{0}
\renewcommand*{\theequation}{S\arabic{equation}}

\begin{center}
{\large \bf Supplemental information:\\
Acceleration and deceleration of quantum dynamics based on inter-trajectory travel with fast-forward scaling theory}\\ \vspace{0.4cm}

{Shumpei Masuda$^{1}$, Jacob Koenig$^{2}$ and Gary A. Steele$^{2}$}\\
\vspace{0.2cm}
$^{1}$ {\it Research Center for Emerging Computing Technologies (RCECT), National Institute of Advanced Industrial Science and Technology (AIST), 1-1-1, Umezono, Tsukuba, Ibaraki 305-8568, Japan}\\
$^{2}$ {\it Kavli Institute of Nanoscience, Delft University of Technology,\\
 Lorentzweg 1, 2628 CJ, Delft, The Netherlands}\\
\vspace{0.2cm}
%$^\ast$emai: shumpei.masuda@aist.go.jp, j.d.koenig@tudelft.nl
\end{center}

\section*{S1 Realization with superconducting qubits}
\label{superconducting qubits}
We consider the case of two superconducting transmon qubits with a fixed capacitive coupling $g$ far less than the resonance frequency of either qubit. 
The Hamiltonian of the system can be written as 
\begin{eqnarray}
H &=& \omega_1(t) a^\dagger a + \frac{\alpha_1}{2}a^\dagger a^\dagger a a + \omega_2 b^\dagger b + \frac{\alpha_2}{2}b^\dagger b^\dagger b b + g(a^\dagger b +  b^\dagger a),
\end{eqnarray}
where $\alpha_i$ is the anharmonicity parameter of qubit $i$ and we work in units where $\hbar=1$. Each qubit's resonance frequency and the coupling between them may be written as 
\begin{eqnarray}
\omega_i=\sqrt{8E_{Ji}E_{Ci}}-E_{Ci}
\end{eqnarray}
\begin{eqnarray}
g=\frac{E_{Cc}}{\sqrt{2}}(\frac{E_{J1}E_{J2}}{E_{C1}E_{C2}})^{1/4}
\end{eqnarray}
provided $E_{Ji}  \gg E_{Ci}$, where $E_{Ji}$ and $E_{Ci}\approx-\alpha_i$ are the Josephson and charging energies of qubit $i$ respectively, with $E_{Cc}$ the charging energy of the coupling capacitor \cite{Koch2007S, Didier2018S}. For sufficiently large anharmonicity, we may truncate the above Hamiltonian and obtain equation (1). To realize the scheme outlined in the main text, the most straightforward approach is to use one tunable-frequency qubit with resonance frequency $\omega_1(t)$ and one fixed-frequency qubit with $\omega_2$. We consider an asymmetric transmon for which the two Josephson junctions which comprise its SQUID loop have different Josephson energies. The frequency of qubit $1$ is tunable by varying the applied flux $\Phi$ through the loop, given that 
\begin{eqnarray}
E_{J1}(\Phi)=E_{J1}^{\rm max}\cos{\Big{(}\frac{\pi\Phi}{\Phi_0}\Big{)}\sqrt{1+d^2\tan^2{\Big{(}\frac{\pi\Phi}{\Phi_0}\Big{)}}}},
\end{eqnarray}
where $E_{J1}^{\rm max}$ is the total, maximum Josephson energy of the loop, $\Phi_0$ is the magnetic flux quantum, and $d$ is a measure of the junction asymmetry \cite{Hutchings2017S}. The applied flux $\Phi$ may be varied in time to satisfy the requirements on the time-dependent frequency. As an example, when considering
the accelerated dynamics in Fig.~\ref{system_10_4_20}e and the decelerated dynamics in Fig.~\ref{VFF_10_25_20}b, the flux should be smoothly tuned from $\Phi(0)=0$ to $\Phi(T_{F})=0.5\Phi_0$ in a time $T_{F}=0.9g^{-1}$ or $T_{F}=1.1g^{-1}$.
As one particular example, the accelerated and decelerated dynamics can be closely replicated for a total evolution time on the order of $10^2$ns for $E_{J1}^{\rm max}=$ 30GHz, $E_{J2}=27.7$GHz, $E_{C1,2}=203$MHz, $g=9$MHz, and $d=0.85$, which is comparable to modern gate times with transmon qubits and far shorter than standard relaxation and dephasing times typically on the order of tens of microseconds.

\section*{S2 Single qubit}
\label{Single qubit}
In this section, we show that the dynamics examined in the main text can be emulated by a superconducting transmon qubit under a drive field.
The Hamiltonian of the system is written as 
\begin{eqnarray}
\frac{H}{\hbar}= \omega(t) a^\dagger a+ \frac{\alpha_0}{2}a^\dagger a^\dagger a a+ 2\Omega \cos(\omega_{\rm d}t)(a + a^\dagger) ,
\end{eqnarray}
where $\omega$ and $\alpha_0$ are the angular frequency and anharmonicity parameter of the transmon, and $\Omega$ and $\omega_{\rm d}$ are the Rabi frequency and angular frequency of the drive field. Now we move to a rotating frame with frequency $\omega_d$ and use the rotating wave approximation (RWA) to obtain
\begin{eqnarray}
\frac{H_{\rm RWA}}{\hbar}=\Delta(t) a^\dagger a+ \frac{\alpha_0}{2}a^\dagger a^\dagger a a+ \Omega (a + a^\dagger),
\end{eqnarray}
where $\Delta(t) = \omega(t) - \omega_{\rm d}$.
We assume that $|\Delta|, \Omega \ll | \alpha_0 |$.
Then the system can be approximated as a two level system for which the Hamiltonian is represented as
\begin{eqnarray}
H_{\rm RWA}=\frac{\hbar \Delta}{2}  \sigma_z + {\hbar \Omega} \sigma_x,
\end{eqnarray}
where we shifted the origin of the energy by $\hbar\Delta/2$. 
This is effectively the same as the system for which the dynamics is governed by equation~(\ref{SE_ref_10_24_20}).
Therefore, the accelerated and decelerated dynamics examined in the main text can be emulated by this single qubit system under a drive, although the detuning should be sufficiently smaller than $\alpha_0$.

\section*{S3 Role of additional phase}
\label{Role of additional phase}
We discuss the role of additional phase by showing how the relative phase between $\phi_m$ and $\phi_l$ influences the time dependence of the population.
Using equation~(\ref{SE_ref_10_24_20}), the time derivative of the population $|\phi_m|^2$ can be written as
\begin{eqnarray}
\frac{d}{dt}|\phi_m|^2 = -ig\phi_m^\ast \phi_l  + ig\phi_m \phi_l^\ast.
\label{dpop_2_26_21}
\end{eqnarray}
Equation~(\ref{dpop_2_26_21}) can be rewritten as 
\begin{eqnarray}
\frac{d}{dt}|\phi_m|^2 = -2g\tilde\phi_m \tilde\phi_l  \sin(\theta_m - \theta_l),
\label{dpop_2_26_21v2}
\end{eqnarray}
where $\tilde\phi_m$ and $\theta_m$ are the intensity and the phase of $\phi_m$, that is,
$\phi_m=\tilde\phi_m e^{i\theta_m}$ where $\tilde\phi_m, \theta_m \in R$.

Thus, it is seen that the relative phase $\theta_m - \theta_l$ affects the rate of change of the population as well as the intensity $\tilde\phi_m$ and coupling $g$.
The intensity of the wave function and the coupling of the accelerated or decelerated dynamics are the same as in the reference dynamics in our formalism. 
Therefore, deformation of the phase via additional phase is required for acceleration and deceleration.

\section*{S4 Tunable coupling}
\label{Tunable coupling}
The fast-forward scaling theory can be extended straightforwardly to the case in which the coupling strength is tunable. 
The reference dynamics develops under time-dependent $g$ and $\omega_m$.
The Schr\"{o}dinger equation is represented by
\begin{eqnarray}
i\frac{d}{dt}\phi_{m}(t) = g(t) \phi_{l}(t) + \omega_m(t) \phi_{m}(t).
\label{SE_ref_1_31_21}
\end{eqnarray}

We extend the formalism to obtain $g$ and $\omega_m$ which accelerate, decelerate, or reverse the system evolution relative to the reference dynamics.
If $g$ and $\omega_m$ can be perfectly controlled, we can use a trivial scaling property as explained later. 
However, the controllability of the parameters is limited in speed and range by device parameters and control hardware. 
Therefore, it will be meaningful to develop fast-forward scaling theory also for the case with a tunable coupling as the theory would then provide various ways to generate a target state. 

We assume that the wave function, $ \phi_m^{\rm FF}(t)$, of the accelerated, decelerated, or reversed  dynamics has the same form as equation~(\ref{phiFF_1_31_21}).
We assume that $\phi_m^{\rm FF}$ is a solution of the Schr\"{o}dinger equation:
\begin{eqnarray}
i\frac{d}{dt}\phi_m^{\rm FF}(t) = g_{\rm FF}(t) \phi_{l}^{\rm FF}(t) + \omega_m^{\rm FF}(t) \phi_m^{\rm FF}(t).
\label{SEFF_1_31_21_v2}
\end{eqnarray}
In the same manner as used for equations (\ref{f_10_10_20}) and (\ref{V_10_10_20}), we can obtain two equations:
\begin{eqnarray}
\alpha(t) g(t) {\rm Im} [ \phi_m^\ast \phi_{l}] = g^{\rm FF}(t) {\rm Im}\{ \phi_m^\ast \phi_l \exp[i(f_l-f_m)]\} \nonumber\\
\label{f_1_31_21}
\end{eqnarray}
and  
\begin{eqnarray}
\omega_m^{\rm FF}(t) &=& {\rm Re} \Big\{ \frac{\phi_l}{\phi_m} \Big{(}\alpha(t) g(t)- g^{\rm FF}(t)\exp[i(f_l-f_m)]  \Big{)} \Big\}\nonumber\\
&& + \alpha(t) \omega_m(\Lambda(t)) - \frac{df_m}{dt},
\label{V_1_31_21}
\end{eqnarray}
where $l\ne m$, and $\phi_{m(l)}$ and $f_{m(l)}$ abbreviate $\phi_{m(l)}(\Lambda(t))$ and $f_{m(l)}(t)$, respectively.

Equations~(\ref{f_1_31_21}) and (\ref{V_1_31_21}) have a trivial solution $g^{\rm FF}(t)=\alpha(t) g(t)$, $\omega_m^{\rm FF}(t)=\alpha(t) \omega_m(\Lambda(t))$ and $f_m(t)=0$. 
The corresponding dynamics is simply scaled with respect to time without any additional phase.
However, in general, $g^{\rm FF}(t)$ can be chosen to be different from $\alpha(t) g(t)$. 
Thus, equations~(\ref{f_1_31_21}) and (\ref{V_1_31_21}), which encompass the simply scaled dynamics, 
provide us with various choices for the time dependence of the control parameters for acceleration, deceleration, or reversal. 

\section*{S5 Gaps between trajectories}
\label{Gaps between trajectories}
In order to examine the mechanism by which the gaps between trajectories manifest, we rewrite equation~(\ref{f_10_10_20}) as
\begin{eqnarray}
\alpha(t) b(t) = b(t) \cos(f_l(t)-f_m(t)) + a(t) \sin(f_l(t)-f_m(t)),
\label{fver2_1_30_21}
\end{eqnarray}
where $a(t)={\rm Re}[\phi_m^\ast \phi_{l}]$ and $b(t)={\rm Im}[\phi_m^\ast \phi_{l}]$, and 
$\phi_{m(l)}$ abbreviates $\phi_{m(l)}(\Lambda(t))$ in this section.
Equation~(\ref{fver2_1_30_21}) can be rewritten as
\begin{eqnarray}
\frac{\alpha(t) b(t)}{r(t)} = \sin(f_l(t)-f_m(t)+\theta(t)),
\label{fver3_1_30_21}
\end{eqnarray}
where $r(t)=\sqrt{a(t)^2 + b(t)^2}$ and $\tan\theta(t) = b(t)/a(t)$. 
Equation~(\ref{fver3_1_30_21}) has at most two solutions of $f_l(t)-f_m(t)$ for $-\pi \le f_l(t)-f_m(t) < \pi$.
When $\phi_m^\ast \phi_{l}$ is purely imaginary $a(t)=0$, $b(t)\ne 0$, and $\alpha(t)=1$, two solutions are degenerate at $f_l-f_m=0$.  

\begin{figure}
\begin{center}
\includegraphics[width=15.0cm]{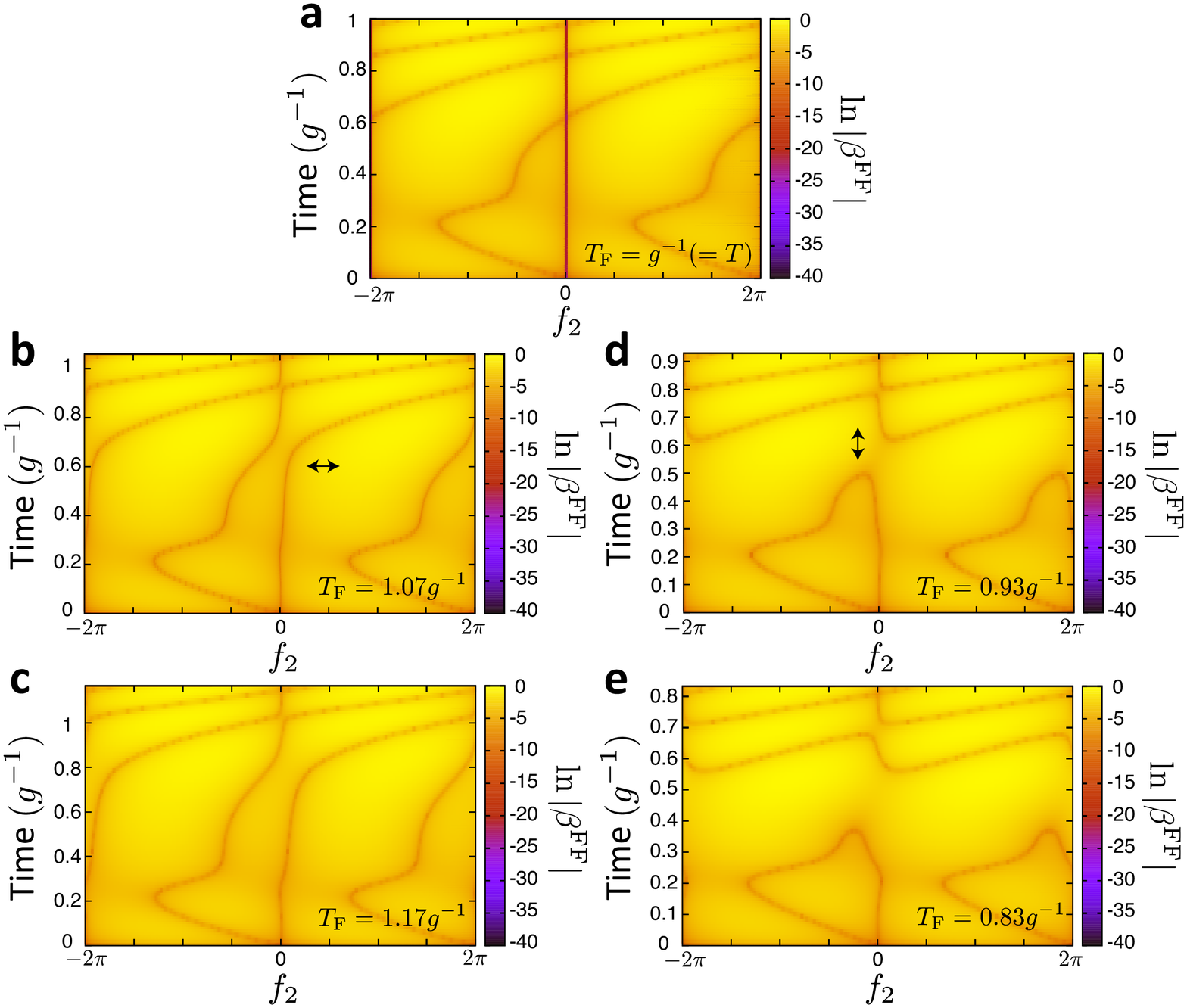}
\end{center}
\caption{{\bf Gap opening between speed-controlled trajectories.}
{\bf a-e,} $\ln|\beta^{\rm FF}|$ as a function of $f_2$ and $t$ for $T_{\rm F}$ indicated in the panels.
Other parameters used are the same as in Fig.~\ref{system_10_4_20}b.
On the brown lines, $f_2=0$ and $f_2=\pm 2\pi$, in {\bf a}, $\beta^{\rm FF}(t,f_2)=0$ ($\ln|\beta^{\rm FF}(t,f_2)|=-\infty$). These lines correspond to the reference dynamics.
{\bf b,c} correspond to deceleration, and {\bf d,e} to acceleration.
The arrows in {\bf b} and {\bf d} represent the direction in which the gap between the trajectories opens.
}
\label{phase_log_com_10_10_20}
\end{figure}

Figure~\ref{phase_log_com_10_10_20} shows $\ln|\beta^{\rm FF}|$ for various values of $T_{\rm F}$ for $f_1(t)=0$.
In Fig.~\ref{phase_log_com_10_10_20}a for $T_{\rm F}=g^{-1}$, there is a trivial trajectory at $f_2=0$ given that $\alpha(t)=1$. This trajectory corresponds to the reference dynamics.
The intersections of the trajectories at $f_2=0$ in Fig.~\ref{phase_log_com_10_10_20}a correspond to the degeneration points where $\phi_m^\ast \phi_{l}$ is purely imaginary. 
The intersections are disconnected and the gap between the trajectories opens in Fig.~\ref{phase_log_com_10_10_20}b--e where $\alpha(t)\ne 1$ for $0<t<T_{\rm F}$.
The gap between the trajectories opens horizontally for the decelerated dynamics as seen in Fig.~\ref{phase_log_com_10_10_20}b and \ref{phase_log_com_10_10_20}c which correspond to $\alpha(t)<1$. 
On the other hand, the gap between the trajectories opens vertically for the acceleration as seen in Fig.~\ref{phase_log_com_10_10_20}d and \ref{phase_log_com_10_10_20}e which correspond to $\alpha(t)>1$.
This is because there are two solutions of $f_2(t)$ for $\alpha(t)<1$, while there is no solution of $f_2(t)$ for $\alpha(t)>1$, when $\phi_m^\ast \phi_{l}$ is purely imaginary.

\section*{S6 Shift of $\omega$}
\label{Relative phase}
We show that differences between $\omega_1(t)$ and  $\omega_2(t)$ give rise to changes in the overall phase of the wave function.
We assume that $\phi_m$ satisfies equation~(\ref{SE_ref_10_24_20}).
Now, we introduce $\bar\phi_m(t)$ defined by
\begin{eqnarray}
\bar\phi_m(t) = \phi_m(t) e^{i\theta(t)},
\end{eqnarray}

where $\theta(t)$ is independent of $m$.
Using equation~(\ref{SE_ref_10_24_20}), we obtain 
\begin{eqnarray}
i\frac{d}{dt}\bar\phi_{m}(t) = g \bar\phi_{l}(t) + \big{(}\omega_m(t) - \dot\theta(t)\big{)} \bar\phi_{m}(t),
\end{eqnarray}
This result represents that the wave function simply acquires additional global phase when $\omega_1(t)$ and $\omega_2(t)$ are shifted by the same amount, $- \dot\theta(t)$.

\end{document}